# Innovative Technologies for Optical and Infrared Astronomy

Colin R. Cunningham[*a], Christopher J. Evans[a], Frank Molster[b], Sarah Kendrew[c], Matthew A. Kenworthy[b] and Frans Snik[b]

[a] UK Astronomy Technology Centre, Royal Observatory Edinburgh, Blackford Hill, Edinburgh, EH9 3HJ, UK
[b] Leiden Observatory, Leiden University, NL-2300RA Leiden, The Netherlands
[c] Max-Planck-Institut für Astronomie, Königstuhl 17, 69117 Heidelberg, Germany

**ABSTRACT**

Advances in astronomy are often enabled by adoption of new technology. In some instances this is where the technology has been invented specifically for astronomy, but more usually it is adopted from another scientific or industrial area of application. The adoption of new technology typically occurs via one of two processes. The more usual is *incremental* progress by a series of small improvements, but occasionally this process is *disruptive*, where a new technology completely replaces an older one. One of the activities of the OPTICON Key Technology Network over the past few years has been a technology forecasting exercise. Here we report on a recent event which focused on the more radical, potentially disruptive technologies for ground-based, optical and infrared astronomy.

## 1. INTRODUCTION

The Key Technology Network (KTN) is one of the networking activities undertaken as part of the OPTical Infrared Co-Ordination Network (OPTICON), funded by the EU Framework 6 and 7 programmes. The objectives of the KTN include: identification of future key technologies which would advance European astronomy, to encourage collaborative technology development projects within Europe, and to facilitate discussions on future technology development. An introduction to the KTN and a roadmap for future technology developments for ground-based, optical and infrared astronomy was presented by Cunningham et al[1]. Here we revisit this roadmap, paying particular attention to potentially 'disruptive' technologies which could significantly affect the design and operation of future telescopes and their instrumentation. The historic background to disruptive technology in astronomy is explored more fully by Cunningham[2]. Some of the ideas regarding free-flying telescope modules came from the Techbreak technology foresight programme run by the European Science Foundation for the European Space Agency.

**1.1 Disruptive Technology**

The concept of disruptive technology was used to describe the situation in commercial markets where a new technology completely changes a given market (Christensen[3]). For example, when vinyl records were replaced by compact discs, the basic nature of music retail did not change – this was essentially an incremental (albeit large) change in the market. However, with the advent of personal digital devices (such as the iPod), the process of distributing and retailing music changed fundamentally, with on-line downloads significantly displacing sales of physical music, with the resulting closure of traditional distributions systems such as music shops. Similar events can be seen in the history of astronomy, such as the introduction of electronic array detectors in the late 1970s, displacing photographic plates and their associated analysis techniques.

With limited resources available for research and development programmes, it is prudent to consider which emerging technologies and techniques have the potential to be disruptive in the future. Anticipating the impact of developments ahead of time is a challenging task, but in April 2012 we had a workshop in Marseille to serve as both a 'brain-storming' and 'horizon-scanning' exercise. We adopted the methodology used by Cunningham et al[1], employing the 'T-Plan' roadmapping methods from Phaal et al[4], which delineates the process of technology development into three stages: market, product, and technology. The analogies for astronomy in this scenario are the science challenges and drivers of the future, the features needed to achieve these (of telescopes, instrumentation, and analysis), and the technological components/solutions needed to enable them.

---

[*] colin.cunningham@stfc.ac.uk

### 1.2 Technology Readiness Levels

In consideration of the different stages of technology development and adoption, we employ the concept of Technology Readiness Levels (TRLs), as used by the National Aeronautics and Space Administration for technology in space missions. In brief, there are three main phases in the TRL scale:

- TRL 1-3: Basic technology research through to demonstration of feasibility;
- TRL 4-6: Further development through to full demonstration;
- TRL 7-9: System-level development through to successful operational use.

### 1.3 TRIZ – a methodology for innovation

Our approach for exploring innovative technology for astronomy has been influenced by the TRIZ movement, started by Genrich Altshuller[5]. TRIZ is a methodology designed to systematize the problem solving process. One technique to help break through to the next level of solution is to imagine the ideal outcome, which often can help get rid of a technological constraint. A good example is shown by the evolution of control in aircraft, where there is a progression from mechanical control of wing surfaces, through hydraulics to 'fly by wire', but perhaps ultimately getting rid of the discrete surfaces (flaps, aerilons) and making the whole wing a 'smart structure' that can change its shape. Here the 'ideal solution' completely removes the need for the previous technology.

*An ideal technological system is a system which does not itself exist as a physical entity but the function that it delivers is still fully performed*

We will see later that imagining that we have a 'magic wand' that could conjure up an ideal solution could suggest some ways to revolutionise astronomical telescopes and instrumentation: a telescope without a structure or adaptive optics without deformable mirrors. An interesting TRIZ concept that can be exposed by such a fantasy is that of 'contradictions'. What is the barrier between what we know how to do now and what we would need to approach the ideal solution? How can we break through that contradiction? Usually it cannot be done by extrapolation from where we are now, but needs a leap into radical new technologies. We will see later that this may well be needed if we ever to build bigger telescopes in space.

## 2. 'GRAND CHALLENGES' FOR GROUND-BASED ASTRONOMY BEYOND 2025

To provide a focus for discussion of future technologies, and the factors influencing their design and requirements, we solicited input from the workshop participants on future *scientific* challenges in astronomy that will require instrument and/or telescope capabilities beyond those currently in use, or in advanced stages of development. Of course, it is difficult to disentangle the scientific motivations from the features of potential instruments/telescopes completely. This is particularly true in the case of improvements in sensitivity, in which all scientific frontiers would benefit from increased performance. To provide some context to the following discussions, we briefly summarise three examples which will no doubt shape future developments:

- **Spectroscopy of planetary atmospheres:** The study of exo-planets is one of the fastest growing areas of modern astronomy, with detection via transits, radial-velocity monitoring (with high-resolution spectroscopy), and high-contrast imaging. Spectroscopy of planetary atmospheres is even more in its infancy, limited at present to some of the more extreme companions (e.g. Janson et al[6]). One of the key motivations for the next generation of ground-based telescopes, the Extremely Large Telescopes (ELTs), is for direct, high-contrast imaging of planets that are lower in mass and closer to their parent stars than possible at present. This potentially opens-up the 'habitable zone' of rocky planets for more detailed study. However, even with the capabilities of the ELTs, it is unlikely that spectroscopy of 'Earth-like' planets will be within their grasp – requiring yet larger telescopes/finer resolution/greater contrast.

- **Optical imaging of stellar populations:** In the coming decade we can expect to see the end of operations with the *Hubble Space Telescope (HST)*. For over two decades we have had access to high-quality imaging in the optical with a stable angular resolution of better than 0.1 arcsecs. A whole generation of astronomers has become accustomed to this sort of performance across a wide range of astronomical topics, yet there is no obvious successor at short visible wavelengths. Studies of resolved stellar populations in external galaxies are one area in particular which benefit greatly from optical imaging – near-IR photometry (even with adaptive optics) does not provide the same leverage on star-formation histories, metallicities and so on. Lucky imaging (Law et al[7]) is likely the best near- to medium-term solution for selected applications, but a much more disruptive development would be ground-based, 'wide-field' (1-2 arcmin) diffraction-limited imaging in the optical via adaptive optics.

- **Redshift surveys in the era of ELTs:** There are a number of studies underway for the next generation of optical and near-IR multi-object spectrographs (MOS), including.4MOST, BigBOSS, DESpec, MOONS, SuMIRe, and WEAVE (see de Jong et al[8], Mostek et al[9], Cirasuolo et al[10], Dalton et al[12]). With a mixture of science cases, they are largely motivated for observations of large numbers of galaxies (to study dark energy and large-scale structure), as well as follow-up of Galactic targets observed with the *Gaia* satellite. With current technologies, this next generation of MOS instruments will likely have a multiplex (number of sources observed in parallel) of a few thousand per observation on the sky. If demand for spectroscopic follow-up continues to grow in the coming two decades (e.g. full catalogues from LSST, VISTA, etc), we can anticipate even larger surveys, which would require new techniques. The science in this case is somewhat incremental, but the technologies required to enable it are more of a step-change from what is used at present. Moreover, if working in the near-IR, sensitivity can also be boosted via 'collecting the right photons', that is improved suppression or subtraction of the OH sky lines.

Two directions were then considered: a 'top-down' approach of the technology required to advance studies of exo-planets in the era of ELTs (Section 3), and a 'bottom-up' approach of discussing what novel components are under development that could fundamentally change the way that astronomy is done (Section 4).

## 3. EXO-PLANET STUDIES IN THE 2030s

One of the long-term ambitions for the ELTs has been direct imaging of exo-planets, with studies (and associated technology development plans) for instruments such as EPICS for the E-ELT (Vérinaud et al[12]). There are still challenges to be met to satisfy the demanding requirements of high-contrast, diffraction-limited imaging with the ELTs, both in terms of ultimate sensitivity, image quality, and our characterisation of the atmosphere.

### 3.1 Limiting factors to imaging techniques

The star-to-planet contrast for rocky planets in the habitable zone at optical/IR wavelengths is of the order $10^9$ or more, beyond the anticipated performance of the next generation of planet-finding instruments (VLT-SPHERE and Gemini Planet Imager). Direct imaging of rocky planets with the E-ELT will require a combination of extreme AO, advanced coronography, and sensitive imaging-polarimetry in the optical (Keller et al[13]). Longer IR wavelengths will be less effective because of the coarser inner working angle (scaling with $\lambda/D$). Sensitive polarimetry helps to deliver the enhanced contrast as the atmospheres of rocky planets are expected to be linearly polarized by tens of percent (as demonstrated by our blue sky).

One of the challenges for EPICS to directly image a rocky planet in the habitable zone is the construction and control of a deformable mirror (DM) with around 40,000 actuators at a rate of up to 3 kHz. Current manufacturing methods are barely able to furnish flawless mirrors with a thousand actuators with sufficient stroke, and this technology does not scale easily to DMs with considerably more actuators. Novel technology in this area would be truly disruptive to exo-planet research as it would open a new part of discovery space. Furthermore, even extrapolating Moore's law to beyond 2020, conventional algorithms for AO control cannot be reasonably implemented for such a DM, so novel, advanced algorithms are being developed.

### 3.2 Searching for biomarkers with spectroscopy

If rocky planets are detected in the habitable zone around parent stars, there will be a huge desire to characterise their atmospheres via spectroscopy. Observations of transiting Earth-like planets may provide detections of biomarkers, but it seems likely that direct-detection methods will remain the more favourable approach, although this requires excellent suppression of the light from the central star due to the contribution to the noise from the halo of speckles. This suppression can be localised (see Trauger and Traub[14]) as the expected separation of the targets will generally be well known from radial velocity and/or transit observations.

Spectro-polarimetry is potentially a powerful tool as it can provide unambiguous information on the presence of water clouds or oceans (Sterzik et al[15]) and oxygen (Stam[16]). However, polarimetric accuracy is just as important as sensitivity to quantify these signatures robustly (Snik and Keller[17]). This means that contributions to the polarisation from the telescope and the instrument need to be well understood or circumvented, for example by implementing polarisation optics within the telescope (in the same way that AO is being integrated into modern telescopes). Furthermore, spectro-polarimetric IFUs could be used for additional speckle rejection (Rodenhuis et al[18]).

A convincing sign of (non-intelligent) life that could be observed remotely is the presence of homochirality in complex molecules such as amino acids and sugars, with the imbalance between one handedness and the other thought to arise from biological processes. Homochirality should be detectable via circular polarisation (Sparks et al[19]), but the signal from a planet covered with primitive life is likely (at least) an order of magnitude smaller than the degree of linear polarisation from atmospheric scattering. Even larger telescopes (with a very good control of polarization cross-talk) will therefore be required to answer the biggest question of all: 'Is there other life out there?' It is highly likely that detection of such biomarkers will only be possible with large space telescopes.

### 3.3 'Predictive Adaptive Optics'

Another limiting factor in ground based high-contrast AO observations is the wavefront error introduced by the delay between measuring the wavefront and sending an appropriate correction to the DM (Guyon[20]; Poyneer and Macintosh[21]). When the atmospheric turbulence is dominated by a single wind flow above the telescope, it is not possible to easily predict what the turbulence will look like along the edge of the telescope pupil facing the wind. An additional complication is the significant size of the secondary mirror, which provides an additional leading edge of turbulence that is not measured. Both of these effects lead to a limit in wavefront sensing contrast of $10^{-5}$ at 1 to 2 $\lambda/D$ (e.g. Guyon[22]). The effects of this delay can be mitigated partially by using multiple natural guide stars (NGS) surrounding the scientific field-of-view and employing tomographic reconstruction of the atmosphere above the telescope, but this comes at a cost of focal-plane complexity and problems in integration with other scientific instruments.

Predicting the next few milliseconds of turbulence given previous telemetry would be an extremely useful breakthrough, and several attempts at predictive wavefront sensing have been made (e.g. Poyneer and Veran[23]). An alternative method to achieve this is to add a ring of smaller (0.2 to 1.0m) telescopes around the perimeter of the ELT dome at a distance of several tens of metres. These 40 or 50 small telescopes would look at a bright NGS near to the scientific field, and provide additional sensing data to a tomographic reconstructor. Ideally, the telescopes on the windward edge of the main aperture would provide wavefront sensor telemetry that would be fed into the ELT AO reconstructor on the telescope at the time. By selecting bright guide stars within two or three degrees of the science field, smaller telescopes could be employed.

Such 'helper telescopes' continuously recording atmospheric data using simple off-the-shelf telescopes – in a similar fashion to the DIMM/MASS seeing-measurement telescopes already in use today – could be the key to delivering the high-performance AO correction required for the next generation of large observatories

### 3.4 Improved tomography from 'helper telescopes'

An even more general approach could result in improvements to wide-field AO methods by providing tomography over a larger volume of the first 1 km of atmosphere above an ELT. An array of small, deployable telescopes could be distributed around the ELT site, each with a low-order complex amplitude wavefront sensor (such as a dOTF

(differential Optical Transfer Function) sensor; Codona[24]) at their focal plane and a radio connection to a central AO processing/tomography unit. With potentially hundreds of these units, each looks at a different sight-line toward a NGS close to the science target, providing a well-sampled characterisation of the low order aberration terms for the atmosphere above the telescope and through the cylinder of turbulence defined by the telescope aperture and science target.

Each helper telescope is autonomous, transmitting their position, current pointing (i.e. target NGS or nearby bright star), and measured aberrations back to a central receiver. A modular approach (i.e. each unit is identical, but with a unique radio transmitted identifier) will reduce costs, and would enable quick replacement in case of failure. These supporting observations would be used to provide a tomographic reconstruction of the atmosphere up to 1 km above the telescope, providing a comprehensive wide-angle, ground-layer model that could be used for wide-field imaging instruments and their AO systems.

## 4. EMERGING TECHNOLOGIES and TECHNIQUES

The key opportunity for opening new parameter space in optical and infrared astronomy is likely to be through increases in angular resolution and sensitivity by deployment of larger telescopes, adaptive optics and interferometry. In particular, characterisation of exoplanets through direct imaging and spectroscopy needs a combination of high angular resolution to separate the planet from its star, high sensitivity to the light from the planet, but most importantly, high contrast techniques to penetrate the scattered light and speckle from the star. The first requires either a space telescope or extreme adaptive optics, whereas the last needs innovative techniques in wavefront manipulation, for instance to create dark patches where the planet is known to be in the image. Some of these techniques could be transformative.

New techniques in optical and IR astronomy are already being adopted from the rapidly developing field of photonic devices (Cunningham[25]), and it is possible that concepts in metamaterials and diffractive optics may be used to develop novel wavefront correction and manipulation systems, using structures to transform light that are not available in nature.

### 4.1. Photonic Devices

The adoption of photonic devices in astronomy is starting to take off, as demonstrated by several papers at this conference showing on-sky demonstrations of OH suppression fibres and integrated spectrometers (Allington-Smith and Bland-Hawthorn[26]). The thrust of these devices is to improve sensitivity of IR imagers and spectrometers on the ground by OH line suppression, or finding ways of increasing multiplex advantage for multi-object spectrometry (Ellis and Bland-Hawthorn[27]). These developments may result in much cheaper and more practical means to carry out wide field redshift or stellar population surveys from the ground that could result in science breakthroughs – see section 4.6.

### 4.2. Customized optical materials and devices

Novel static liquid crystal structures are being developed for phase manipulation (Snik et al[28]; Mawet et al[29]) to produce selective apodisation in order to enhance contrast in regions of the star-planet PSF where it is known that the planet should be. These are static devices, manufactured by lithographically applying orientation patterns to substrates in order to tailor the liquid crystal 3-D structures to form controlled multi-twist retarders. In themselves, they show great promise for improving the direct detection and characterization of exoplanets, but they also demonstrate a new paradigm for the development of customized optical devices. Liquid crystal spatial light modulators (SLMs) have been used for low performance wavefront correction (Love[30]) and it is also possible to generate micro-lens arrays using liquid crystals, to enable both wavefront sensing and correction to be done with the same device (Arines[31]). Liquid crystal SLMs have the disadvantage of exhibiting chromatic phase correction, but a whole new class of devices that could be applied in astronomy is likely to emerge from the very active field of metamaterial research. Metamaterials are engineered materials that exhibit properties not found in nature, such as negative refractive index that enables ideal lenses not limited in spatial resolution by diffraction and invisibility cloaks. We could imagine that rapid advances in this field may one day enable the ideal wavefront corrector to be built. Using the TRIZ technique (section 1.3) let us imagine what the properties of this ideal wavefront corrector might be:

- Transmissive;
- High density;
- Low absorption;
- High dynamic range;
- Fast;
- Incorporating programmable microlens sections within the field for wavefront sensing.

The last point could be very important, as non-common path aberrations are currently the dominant limiting factor for AO performance in exoplanet imaging. Doing wavefront sensing with the science camera could bring big benefits.

While the obvious application for this ideal wavefront manipulator would be in extreme adaptive optics, it could also be used as a low-speed correcting element for extremely large space telescopes – see section 4.6.

**Contradictions:** Currently, metamaterial devices manufactured by conventional lithography are only available at microwave frequencies, although demonstration spatial light modulators that can manipulate amplitude or phase have been made at THz frequencies (Chan et al[32], and Chen et al[33]), and phase holograms at mid-IR wavelengths (Larouche[34]). In future, it may be possible to use nanofabrication techniques to develop the ideal transmissive corrector that could even incorporate wavefront sensing into the same device, considerably reducing common path errors and subsequent speckle.

The challenge of designing and building broad-band devices at optical or infrared wavelengths would be considerable, even without consideration of how to make active devices. Metamaterial applications in astronomy are currently limited to multilayer metal mesh interference filters as pioneered by Prof. P Ade and used in most of the submm and far IR instruments built to date (Ade et al[35]). These have tailored bandpass and edge cut-off spectral profiles. It is possible to imagine future metamaterial devices that could extend such architectures to enable multiple notch profiles for OH suppression in the near IR, contributing to the ideal solid-state instrument.

### 4.3. Energy Sensitive Detectors

Continuing the contemplation of the ideal astronomical instrument, if we had solid-state wavefront correction and integrated optical elements, what would the ideal detector be to complement these technologies? The dominant detectors for optical and infrared instrumentation are semiconductor devices such as CCDs and multiplexed diode arrays. These have been very successful, and approach the ideal in terms of quantum efficiency, noise, pixel count and spectral response. However, they are limited by the large band-gap of the semiconductors to only measure intensity of the light detected. Superconducting detectors can be used to measure photon arrival time and energy, opening the path towards completely solid state spectroscopy. Previous attempts to build superconducting cameras with low resolution spectroscopy using Superconducting Tunnel Junctions (STJs: Verhoeve et al[36]) or Transition Edge Superconductors (TESs: Romani et al[37]) were limited to very low pixel counts due to the number of readout wires needed. While the SCUBA2 TES camera (now working very well at submm wavelengths on the James Clerk Maxwell Telescope) has over 10,000 pixels, moving to megapixel TES cameras would be very problematic.

Mazin et al[38] have recently reported a 1024 pixel lumped element Microwave Kinetic Inductance Device (MKID) that works at UV to near IR wavelengths, and could be expanded to megapixel devices in the future by exploiting the multifrequncy microwave readout that enables hundreds of pixels to be interrogated on one wire. Spectral resolution is currently low, but in principle could reach $R/\Delta R$ of a few hundred, and so enable a solid-state integral field spectrometer (Figure 1). While we are speculating on the perfect instrument, let us also propose that the 'magic' metamaterial wavefront manipulator also incorporates programmable filtering, including OH suppression, which would enable many science goals to be achived without the need for high spectral resolution in the detector. A challenge with superconducting devices is clearly the cryogenics, with operating temperatures below 100mK being needed, but sub-100mK cryogen-free cooling is now becoming much easier, with compact Adiabatic Demagnetisation Refrigerators (Bartlett et al[39]) and even the possibility of coolers being incorporated into instruments as solid-state units.

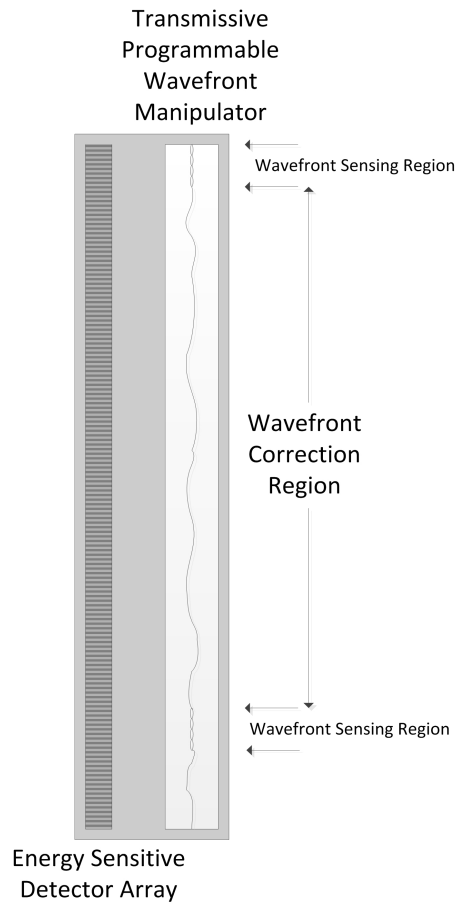

Figure 1. The ideal instrument: wavefront measurement, wavefront correction and imaging spectroscopy in a solid-state module

**4.4. Future multi-object spectrographs**

There is considerable ongoing activity in development of novel multi-object spectrographs. One key area of work is toward OH suppression and 'photonic spectrographs' which should be enabled by some of the components in the previous section and would be transformative in IR spectroscopy. Another active area of research is methods for target selection (e.g. Cochrane et al[40] and Gilbert et al[41]). A push toward larger multiplexes, combined with optimally exploiting the image quality from future AO systems, would likely require techniques such as 'diverse-field spectroscopy' (Murray and Allington-Smith[42]) in which the focal plane is paved with fibres, with switches to select which fibres (i.e. objects) are then coupled to the spectrographs.

If we can really make solid-state spectrometer modules as postulated in section 4.3, then it may be that these could be small enough so that many could be deployed robotically across a wide field to make the ultimate multi-object spectrometer.

**4.5. Real-time calibrations**

A major challenge for achieving the best performance in ground-based optical/IR observations is posed by the Earth's atmosphere. While the effects of atmospheric turbulence can be mitigated by AO, it has become increasingly clear that the ultimate scientific performance of AO-corrected instruments also depends on calibration of the variations in atmospheric transmission, both on short temporal and spatial scales, and as a function of wavelength. Telluric subtraction (the removal of spectral lines originating in the Earth's atmosphere) is usually achieved via observations of standard stars (e.g. Vacca et al[43]), but its success can be limited by variations in airmass, uncertainties arising from the standards themselves (e.g. variability), and characterisation of the instrumental PSF. Indeed, residual errors from telluric subtraction are likely to limit the achievable spectral fidelity of future IR spectrographs (Uttenthaler et al[44]).

Atmospheric transparency can also limit the instrumental performance for imaging. Large-scale imaging surveys such as the Dark Energy Survey (DES; Flaugher[45]), Pan-STARRS (Kaiser et al[46]), and the Large Synoptic Survey Telescope (LSST; Burke et al[47]) require a photometric stability in the optical and NIR of <1% over timescales of years. Even with precise knowledge of the instrumental response and filter bandpasses, Rayleigh scattering and absorption by molecules and Mie scattering by aerosols can cause fast fluctuations in atmospheric transmission affecting the photometric stability of the system.

Real-time atmospheric data can substantially improve the precision of such atmospheric calibrations or corrections, predominantly via monitoring of water vapour using radiometers or GPS-based techniques, for instance in high-resolution IR spectroscopy (Mandell et al[48]; Seifahrt et al[49]), in broad-band near-IR photometry (Burke et al[50]; Blake and Shaw[51]), and in correction for atmospheric dispersion in the mid-IR (Skemer et al[52]). Meanwhile, Burke et al[50] have demonstrated an ambitious scheme where repeated spectroscopic measurements of a catalogue of bright stars can provide the required data when combined with atmospheric transmission models. Improvements have also come from continued advances in molecular line data from the HITRAN database (Rothman et al[53]) and the (free) availability of advanced atmospheric radiative transfer codes.

Some of the areas above could make use of the 'helper telescopes' mentioned in section 3 to improve the processing and calibration of data and, similar to adaptive wavefront sensing, to potentially feed into real time, model-based calibrations.

### 4.6. Even bigger telescopes?

Characterisation of the atmospheres of rocky planets would clearly benefit from even larger telescopes approaching the 100m class on the ground, or perhaps 30m in space, as both the inner working angle and the photon flux from the target planet will be improved; larger telescopes may also open-up imaging of rocky planets at IR wavelengths. The construction of yet larger optical-IR telescopes might be technically feasible (albeit very challenging!), but the costs would likely be prohibitive without novel manufacturing technology or a new telescope design paradigm.

The contradiction here is that currently known techniques to build larger telescopes, such as those developed for the James Webb Space Telescope have so far failed to break the cost paradigm, in that use of segmented mirrors in the *JWST* has not turned out to fulfil the promise of reducing the cost per unit mirror area significantly compared with the *HST*. Maybe the step to larger telescopes of over 30m diameter will need new concepts such as unfolding Fresnel masks (Koechlin et al[54]) or Antoine Labeyrie's Hyper Telescope concept (Labeyrie[55]). Diffractive primary optics solutions such as the photon sieve (Anderson[56]) or the primary objective grating (Dittoscope; Ditto et al[57]) may turn out to be practical alternatives.

Using the TRIZ 'magic wand' approach, let us imagine what an ideal space telescope might look like, especially when combined with the 'ideal instrument' concept we introduced earlier. One way of building lower cost and larger apertures would be to relax the tolerances on the optical surface considerably and then use our 'magic' wavefront transmissive sensor and corrector to correct the errors. Even with our ideal corrector, it may well not be possible to correct all the errors, so the large mirror may need a degree of active control – perhaps using built-in smart actuators such as piezo-fibres. This large mirror could be a free-flying unit and the correction and science detection could be carried out in separate free-flying modules – similar to concepts proposed for X-ray telescopes. The 'ideal instrument' modules could be light and compact, with no mechanical moving parts. The next step then could be to fly a flotilla of such modules so that new modes of operation or wavebands could be employed by swapping modules rather than the usual filter wheels and so on. It may even be possible to maintain the life of the light collector and upgrade the performance of the system by launching new instrument modules as technology progresses.

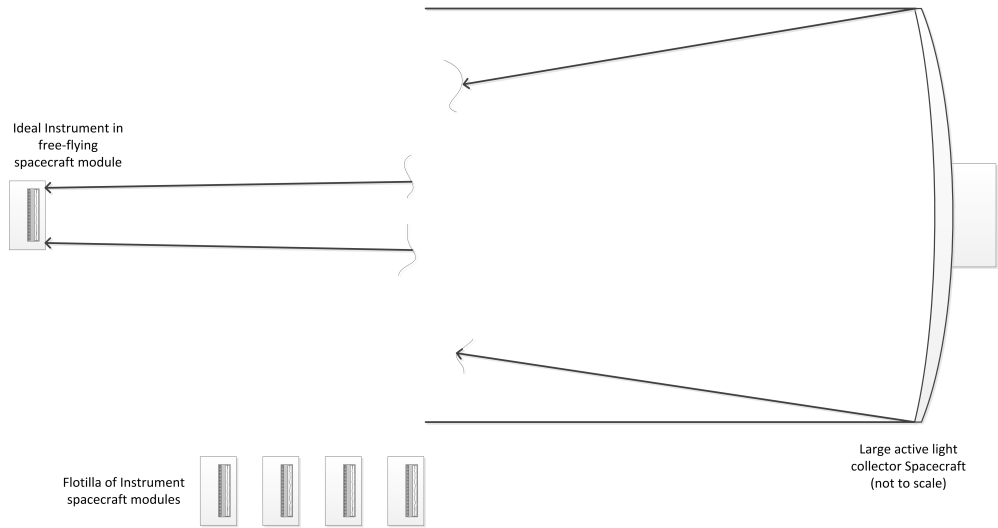

Figure 2. Ideal Space telescope

## 5. OBSTACLES TO DISRUPTIVE TECHNOLOGY

The workshop also considered some of the factors which limit or provide obstacles to early adoption of disruptive technology in the astronomical community. Ultimately, these are all related to the balance of potential cost/risk vs. the gain from innovation.

### 5.1. The development gap

One of the prime obstacles is often to bridge the gap between basic technology development (TRLs 1-3) and adoption at a system level in real instruments/facilities (TRL ≥7). For instance, in the case of novel technology, (industrial) research laboratories will often fund/follow a project through to tests of feasibility (leading to patents and/or publications), but have little incentive to progress it further. Similarly, if technologies are being adopted from other fields, there are often small-scale funds/grants available to develop conceptual demonstrators, but not the more significant funding to develop it to the next stage.

The large overheads to proposals and administration of research funds were noted as a further obstacle, with a 'high risk' development fund suggested as a future line in the ESO instrumentation plan. Some of this type of work is already underway as part of the regular ESO programme, for instance detector development for the GRAVITY instrument (Eisenhauer et al[58]. 2011) and prototyping of key components for the E-ELT. The OPTICON programme has been very successful in funding joint research activities where partners across Europe have developed novel technologies that would have been difficult to fund at national level. Germany has the Reinhart Koselleck funding scheme for exceptionally innovative or higher-risk projects that could be a good model for wider adoption[59].

Where possible we advocate that, for example, the basic technology work carried out under the research activities of the OPTICON programme, are taken forward to higher TRLs via national and/or observatory funding as appropriate. A good example of a project that has been helped to bridge this gap is the development of the CANARY pathfinder experiment of Multi-Object Adaptive Optics on the William Herschel Telescope in La Palma (Gendron et al[60]). The basic components had been demonstrated in the laboratory, but tests of the full system on the sky required more substantial funds from the national research councils.

Many innovations in astronomy come from adoption of new technology that has been developed for other applications, where development budgets can be much higher, for instance the defence industry has been the source of much of our detector technology. A common process is that astronomy adopts a technology from the military such as infrared arrays or adaptive optics, and then pushes the performance to meet demanding requirements, in turn providing improved devices and systems that can be applied in industrial or medical fields – a cycle of innovation. This may turn out to be the route to some of the more fanciful technologies that we have imagined in this paper, such as the transmissive wavefront sensor and manipulator. Exploring common requirements and innovative ideas across a range of diverse applications may well be mutually beneficial.

## 5.2. Demonstrators/visiting instruments

Related to this development gap, our workshop highlighted a general feeling that it is sometimes easier, even for European researchers, to access North American telescopes for on-sky demonstrators (e.g. SWIFT, Tecza et al[61]; FIRST, Huby et al[62]). Upon further discussion, it was noted that such developments do happen on European facilities, e.g., Astralux (Hormuth[63] et al) at Calar Alto, and ExPo (Rodenhuis et al[64]; Canovas[65]) and CANARY (Gendron et al[60]) on the William Herschel Telescope. One factor shaping this view is perhaps just the larger number of small-to-medium sized telescopes in North America, which can be used in the context of technology demonstrators. Further discussion noted a perception of conservatism with regard to ESO facilities in this regard, both in the community in terms of finding funds/making proposals, and from ESO in the available effort/facilities for visiting instruments. Nevertheless, further discussion identified examples of novel, visiting instruments at ESO such as ULTRACAM (Dhillon et al[66]), ULTRASPEC (Dhillon et al[67]), MAD (Marchetti et al[68]), and DAZLE (McMahon et al[69]). There will soon be a full suite of VLT instruments at Paranal, with all potential foci (Cassegrain and Nasmyth) in use. Two options for future access for visiting instruments/technology demonstrators at ESO could be a dedicated telescope at La Silla, or a dedicated visitor focus on the VLT (once older instruments are decommissioned).

## 5.3. Conservatism

It was noted that the funding agencies and, to some extent, the scientific community, can be relatively conservative in their acceptance of risk in new instrumentation. At times this has the perceived effect of stifling innovation and inhibiting areas in which progress could be made more quickly if, for example, 'non-standard' components could be used. Nevertheless, the foundation of the success of observatories such as Paranal has been to deliver robust and effective *facility-class* instruments, which are an integral capability for a decade or more. With such an investment, it is understandable that any risk needs to be well managed and understood. If we are to make large strides toward disruptive technologies in, for example, second generation instruments for the E-ELT, it is evident that we need first to keep making use of smaller telescopes for demonstrators.

## 6. SUMMARY

The OPTICON workshop on disruptive technology in optical and infrared astronomy gathered and generated many ideas on what new technologies could do to transform how astronomy is done, although it should be noted that few were really revolutionary. That is perhaps not surprising, as there are generally only a few of those per century. However, we have noted that the new science and technologies of tailored optical materials and structures could bring about a revolution in telescope and instrument design and performance, even though at present we have no idea how that would come about without resource to the TRIZ 'magic wand'. But don't forget Arthur C Clarke's third law:

*Any sufficiently advanced technology is indistinguishable from magic*

## Acknowledgements

This work was supported by the OPTICON programme, funded from the European Community's Seventh Framework Programme (FP7/2007-2013) under grant agreement number RG226604. We thank the other participants of the KTN Workshop on Disruptive Technologies: Jeremy Allington-Smith, Martin Cullum, John Davies, Kjetil Dohlen, Gert Finger, Roger Haynes, Pierre Kern, David Le Mignant, Ian Parry, William Taylor, and Filippo Zerbi; and Eike Kirke for ideas on space technology contributed via the Techbreak process.